\documentclass[a4paper]{jpconf}
\usepackage{graphicx}
\begin{document}
\title{Chiral Symmetry Breaking in Three Dimensional QED}

\author{Costas Strouthos$^1$, John B. Kogut$^{2,3}$}

\address{$^1$ Department of Mechanical Engineering, 
University of Cyprus, Nicosia 1678, Cyprus} 
\address{$^2$ Department of Energy, Division of High Energy Physics, Washington, DC 20575 USA} 
\address{$^3$ Department of Physics, University of Maryland, College Park, MD 20742, USA} 

\ead{strouthos@ucy.ac.cy, John.Kogut@science.doe.gov}

\begin{abstract}
Over the past few years three dimensional Quantum Electrodynamics (QED$_3$) has attracted a
lot of attention, because it may be an effective theory for the underdoped and non-superconducting
region of the phase diagram of high $T_c$ cuprate compounds.
We present results from lattice simulations of the non-compact version of the theory in order to address
the issue of whether chiral symmetry is spontaneously broken 
when the number of fermion flavours $N_f$ is less than a critical value $N_{fc}$.
Our results provide strong evidence that QED$_3$ is chirally symmetric for  $N_f \geq 1.5$, implying 
that a pseudogap phase separates the superconducting phase from the antiferromagnetic phase.

\end{abstract}

\section{Introduction}
Interest in QED$_3$
has recently been revived by the suggestion that the model may be
an effective theory for the underdoped and non-superconducting region of
the phase diagram of high-$T_c$ superconducting cuprate compounds
\cite{tesanovic.02, herbut}.
In brief, superconductivity in these substances is confined to planes defined by
CuO$_2$ layers, thus motivating a $(2+1)d$ description.
The superconducting order parameter has a $d$-wave symmetry, implying that there
are four nodes in the gap function as the Fermi surface (which
in (2+1)$d$ is a curve) is circumnavigated. At each node the low-energy
quasiparticle excitations obey an approximately linear dispersion relation
with the result that it is possible to rewrite the action
for eight distinct low energy species (spin up and spin down at each of
four nodes) in a relativistically invariant form in terms of $N_f=2$ species
of four-component Dirac spinors. For phenomenologically relevant models
the action for each individual flavor exhibits a spatial anisotropy, a feature
ignored in this paper. If QED$_3$ is a relevant effective theory for cuprates, then
the abstract theoretical problem of
the value of $N_{fc}$ assumes concrete phenomenological reference. If
$N_{fc}>2$, then the theory is chirally broken at zero temperature. On
retranslating from the Dirac spinor basis to the original electron degrees of
freedom, the chiral order parameter is reinterpreted as an order parameter for
spin density waves, whose wavevector gets shorter and shorter as doping is
decreased, until at zero doping the N\'eel antiferromagnetic state is recovered
\cite{herbut}. This picture therefore predicts the existence of a phase boundary
between superconducting (dSC) and antiferromagnetic (AFM) phases at some non-zero doping in
the zero temperature limit. If, on the other hand, $N_{fc}<2$, the chirally
symmetric ground state manifests itself as a tongue of ``pseudogap'' phase
separating dSC from AFM, in which normal Fermi liquid properties may be modified
as a result of a non-perturbative anomalous dimension for the fermion field
\cite{tesanovic.02}. In addition, QED$_3$ has found interesting applications in
the unconventional quantum Hall effect in graphene \cite{gusynin.05}.

The study of quantum field theories in which the ground state shows a sensitivity to the number of
fermion flavours $N_f$ is
intrinsically interesting. 
Apparently, for $N_f > N_{fc}$, the attactive interaction between a fermion and
an antifermion due to photon exchange is overwhelmed by the fermion
screening of the theory's electric charge.
Initial studies of QED$_3$ based on Schwinger Dyson equations (SDEs) using the
photon propagator derived from the leading order $1/N_f$ expansion
suggested that for $N_f$ less than $N_{fc} \simeq3.2$
chiral symmetry is broken \cite{pisarski}. 
Other studies
taking non-trivial vertex corrections into account predicted chiral symmetry
breaking for arbitrary $N_f$ \cite{pennington}. Studies which treat
the vertex consistently in both numerator and denominator of the SDEs
have found $N_{fc}<\infty$, with a value either in agreement with the
original study \cite{maris96},
or slightly higher $N_{fc}\simeq4.3$ \cite{nash}.
An argument based on a thermodynamic inequality
predicted $N_{fc}\leq{3\over2}$ \cite{appelquist}, a result that was later challenged
in \cite{mavromatos}. 
Progress in the direction of gauge covariant solutions for the propagators of QED$_3$
showed that in the Landau gauge a chiral phase transition exists
at $N_{fc} \approx 4$ \cite{maris05}. A gauge invariant determination of $N_{fc}$ based on the
divergence of the chiral susceptibility gives $N_{fc} \approx 2.16$ \cite{tesanovic.03}. It has also been shown that the issue of the gauge dependence of $N_{fc}$ extracted from SDEs
becomes irrelevant if Landau-Khalatnikov-Fradkin transformations are 
taken into account \cite{bashir}.

Recent lattice simulations showed that chiral symmetry is broken for
$N_f=1$, whereas $N_f=2$ appeared chirally symmetric with an upper bound of $10^{-4}$
on the dimensionless condensate \cite{hands2002}.
The principal obstruction to a definitive answer has been
large finite volume effects resulting from the presence of a massless photon in
the spectrum, which prevent a reliable extrapolation to the thermodynamic limit.
Recent lattice simulations of the three-dimensional Thirring model, which may have the same
universal properties as QED$_3$, predicted $N_{fc}=6.6(1)$ \cite{thirring}.
In this paper we present preliminary results from lattice simulations of QED$_3$ on large lattices 
in an effort to detect chiral symmetry breaking for $N_f=0.5,...,2$. 

\section{Lattice Model and Simulations}
We are considering the four-component formulation of QED$_3$ where the Dirac
algebra is represented by the $4 \times 4$ matrices $\gamma_0$, $\gamma_1$
and $\gamma_2$. This formulation preserves parity and gives each spinor
a global $U(2)$ symmetry generated by
$\bf1$, $\gamma_3, \gamma_5$ and $i\gamma_3 \gamma_5$; the full
symmetry is then $U(2N_f)$. If the fermions acquire dynamical mass
the $U(2N_f)$ symmetry is broken spontaneously to $U(N_f) \times U(N_f)$
and $2N_f^2$ Goldstone bosons appear in the particle spectrum.

The action of the lattice model we study is
\begin{eqnarray}
S &=&\frac{\beta}{2} \sum_{x,\mu<\nu} \Theta_{\mu \nu}(x) \Theta_{\mu \nu}(x)
+ \sum_{i=1}^N \sum_{x,x^\prime} {\bar \chi}_i(x) M(x,x^\prime)
\chi_i(x^\prime)
\label{eq:action}\\
\Theta_{\mu \nu}(x) &\equiv& \theta_{x\mu}+\theta_{x+\hat\mu,\nu}
-\theta_{x+\hat\nu,\mu}-\theta_{x\nu}\nonumber\\
M(x,x^\prime) &\equiv&
m\delta_{x,x^\prime}+\frac{1}{2} \sum_\mu\eta_{\mu}(x)
[\delta_{x^\prime,x+\hat \mu} U_{x\mu}
-\delta_{x^\prime,x-\hat \mu} U_{x-\hat \mu,\mu}^\dagger].\nonumber
\end{eqnarray}
This describes interactions between $N$ flavours of
Grassmann-valued staggered fermion fields
$\chi,\bar\chi$ defined on the sites $x$ of a three-dimensional cubic lattice, and
real photon fields $\theta_{x\mu}$ defined on the link between nearest neighbour
sites $x$, $x+\hat\mu$. Since $\Theta^2$ is unbounded from above, eq.(\ref{eq:action})
defines a non-compact formulation of QED; note however that to ensure local gauge
invariance the fermion-photon interaction is encoded via the compact connection
$U_{x\mu}\equiv\exp(i\theta_{x\mu})$, with $U_{x+\hat\mu,-\mu}=U^*_{x\mu}$.
In the fermion kinetic matrix $M$ the Kawamoto-Smit phases
$\eta_\mu(x)=(-1)^{x_1+\cdots+x_{\mu-1}}$
are designed to ensure relativistic covariance in the continuum limit, and $m$ is the bare
fermion mass.

If the physical lattice spacing is denoted $a$, then
in the continuum limit $a\partial\to0$, eq.(\ref{eq:action}) can be shown to be
equivalent up to terms of $O(a^2)$ to
\begin{equation}
S=\sum_{j=1}^{N_f}\bar\psi^j[\gamma_\mu(\partial_\mu+igA_\mu)+m]\psi^j
+{1\over4}F_{\mu\nu}F_{\mu\nu}
\label{eq:Scont}
\end{equation}
ie. to continuum QED in 2+1 euclidean dimensions,
with $\psi,\bar\psi$ describing $N_f$ flavours of four-component Dirac spinor acted
on by 4$\times$4 matrices $\gamma_\mu$, and
$N_f\equiv2N$. The continuum photon field is related to the lattice field via
$\theta_{x\mu}=agA_\mu(x)$, with dimensional coupling strength $g$ given
by $g^2=(a\beta)^{-1}$, and the field strength
$F_{\mu\nu}=\partial_\mu A_\nu-\partial_\mu A_\nu$. The continuum limit is thus
taken when the dimensionless inverse coupling $\beta\to\infty$.


Our numerical simulations were performed using the
standard Hybrid Molecular Dynamics algorithm. We checked the effects of 
lattice discretization on the values of the chiral condensate   
by comparing data extracted from simulations at fixed physical volume $(L/\beta)^3$ and 
fixed physical mass $\beta m$ \cite{strouthos.lat07}.  
The results show that the lattice discretization
effects are small for $N_f>0.5$ at $\beta=0.90$, $m=0.005$ on $54^3$ lattices, 
whereas for $N_f=0.5$ there is an $\sim 8\%$ discrepancy 
between the values of the dimensionless condensate $\beta^2 \langle \bar{\psi} \psi \rangle$ 
at $\beta=0.90$ and $\beta=1.20$. 

\begin{figure}[h]
\begin{minipage}{18pc}
\includegraphics[width=18pc]{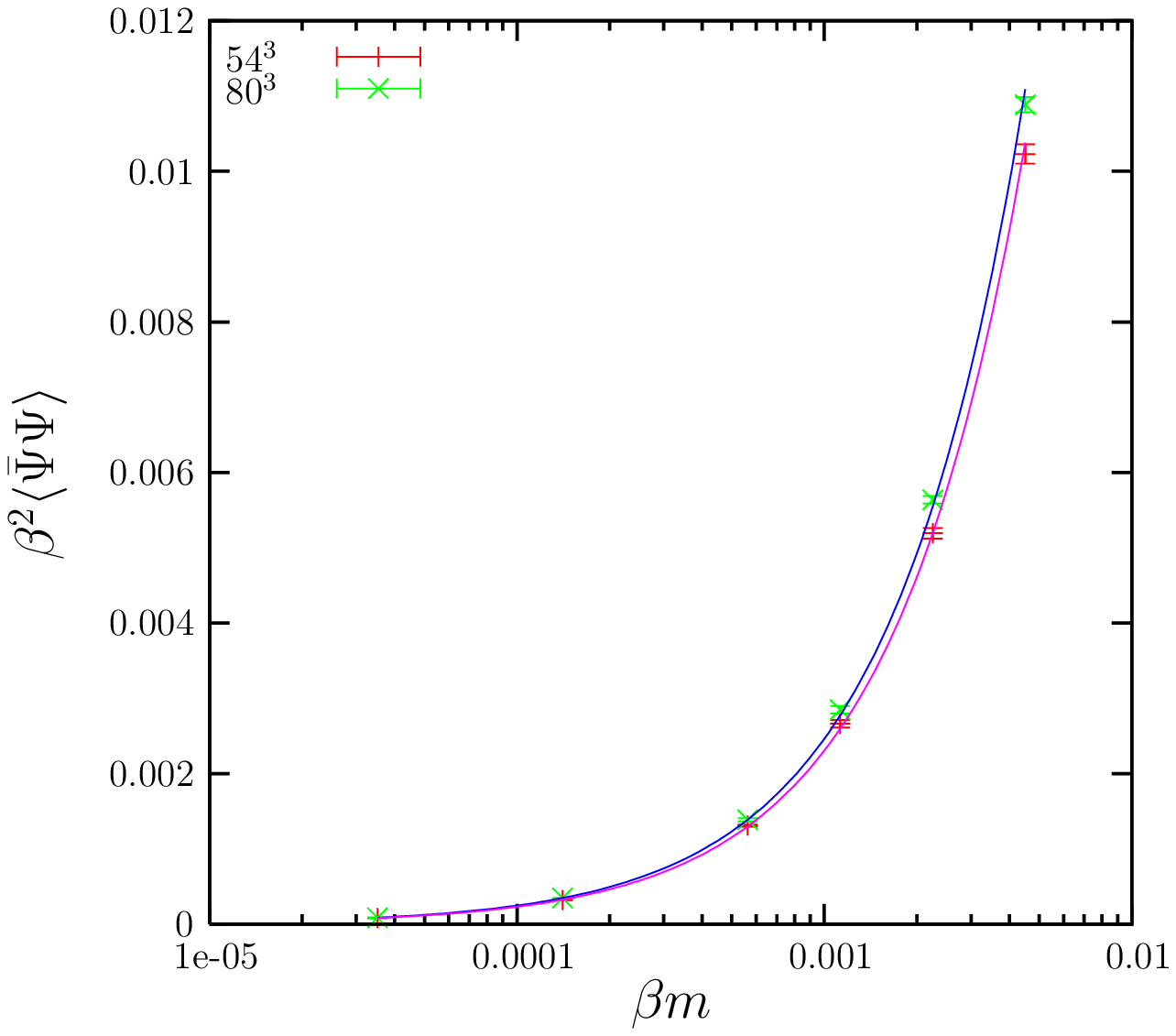}
\caption{\label{fig1} $\beta^2 \langle \bar{\psi} \psi \rangle$ vs $\beta m$ for $N_f=1.5$.}
\end{minipage}\hspace{1pc}%
\begin{minipage}{18pc}
\includegraphics[width=18pc]{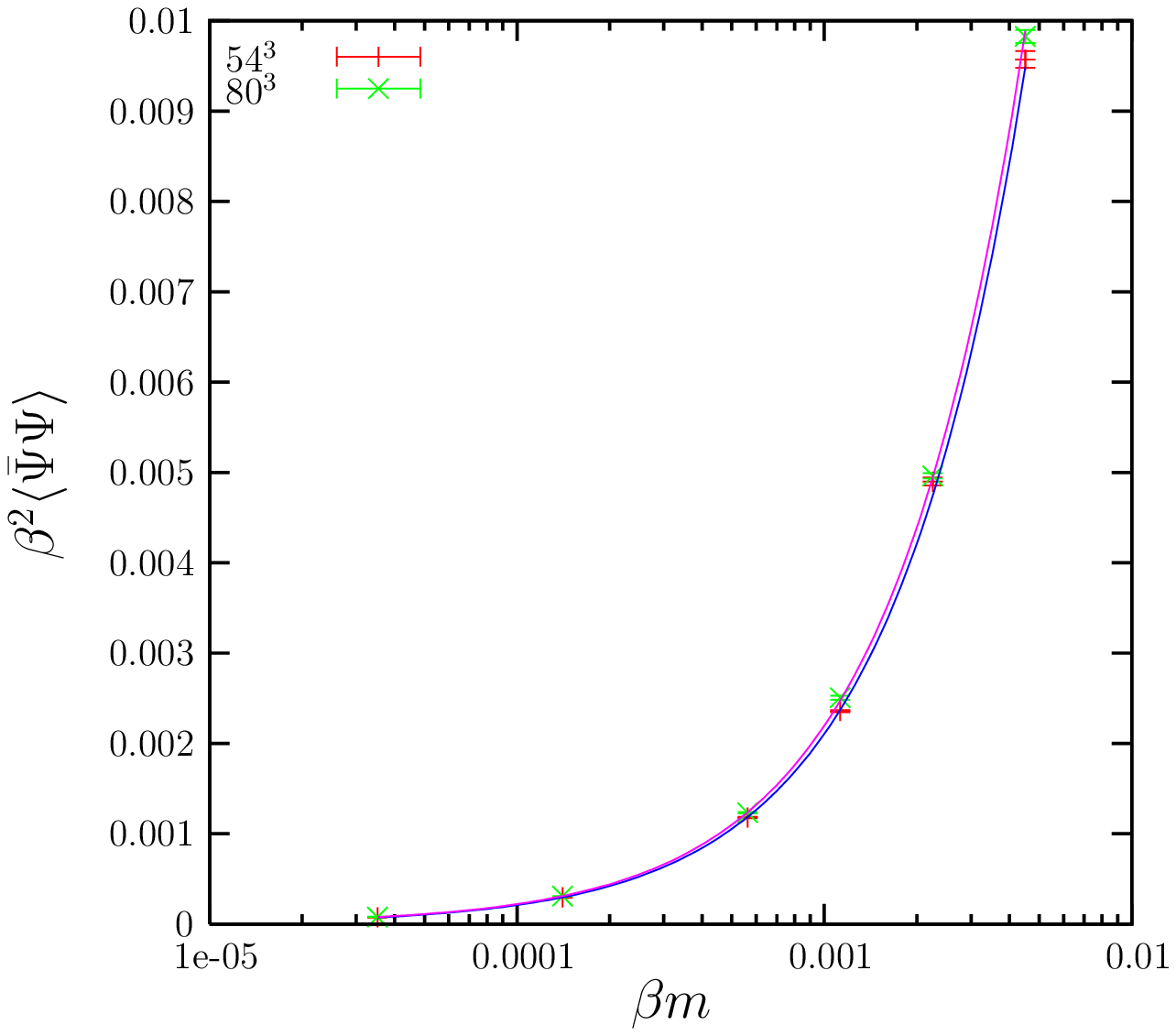}
\caption{\label{fig2} $\beta^2 \langle \bar{\psi} \psi \rangle$ vs $\beta m$ for $N_f=2$.}
\end{minipage}
\end{figure}

We fitted $\beta^2 \langle \bar{\psi} \psi \rangle$
at different values of $\beta m$ for $N_f=1.5$, $N_f=2$ and fixed $\beta=0.90$ to
\begin{equation}
\beta^2 \langle \bar{\psi} \psi \rangle = a_0 + a_1\cdot(\beta m).
\label{eq:fit1}
\end{equation}
The finite size effects are small especially for $N_f=2$ [see Figs. (\ref{fig1}) and (\ref{fig2})]. 
For $N_f=1.5$, $80^3$ we extracted from eq.(\ref{eq:fit1}) $a_0 = -6 \times 10^{-7}$ 
with a statistical error $8 \times 10^{-7}$, 
whereas for $N_f=2$, $80^3$ we extracted $a_0 = 1.5 \times 10^{-6}$ with a statistical error
$10^{-6}$. These results provide evidence that QED$_3$ with $N_f=1.5$ and $N_f=2$ is chirally symmetric
with an accurancy $10^{-6}$. 

\begin{figure}[]
\begin{center}
\includegraphics[width=19pc]{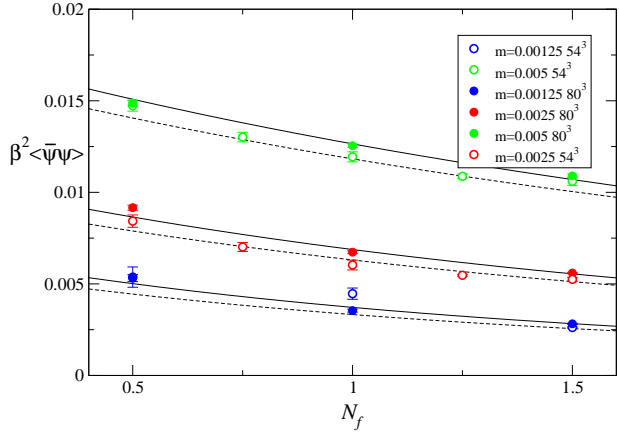}
\caption{\label{fig3} Fits of $ \beta^2 \langle \bar{\Psi} \Psi \rangle$ vs. 
$N_f$ to a finite volume scaling form of the equation of state.
}
\end{center}
\end{figure}

Next, we fitted the values of the dimensionless condensate at different $N_f$,
$m$, and lattice sizes to a renormalization group inspired equation
of state that includes a finite size scaling term \cite{deldebbio}:
\begin{equation}
m=A((\beta-\beta_c)+CL^{-{1\over\nu}}) (\beta^2 \langle\bar{\psi}\psi\rangle)^p+
B (\beta^2 \langle\bar{\psi}\psi\rangle)^\delta ,
\end{equation}
where $p=\delta-1/\beta_m$.
The results extracted from this fit are:
$A=0.0477(38),  B=0.79(2), C=10.7(8), N_{fc}=1.52(6), \delta=1.177(7), p=0.73(2)$.
The data and the fitting functions are shown in Fig. \ref{fig3}.
These results are consistent with a second order phase transition.
 
\section{Summary}
The extrapolations of $\beta^2 \langle \bar{\psi} \psi \rangle$ vs $\beta m$ 
to the chiral limit on lattices with small finite size effects show 
that QED$_3$ with $N_f \geq 1.5$ is chirally symmetric with an accuracy of $O(10^{-6})$, which is in 
agreement with the theoretical prediction of \cite{appelquist}. 
This may imply that the $N_f=2$ chirally symmetric ground state manifests itself in the phase diagram of the 
cuprate compounds 
as a tongue of ``pseudogap'' phase separating the superconducting from antiferromagnetic phases, 
in which normal Fermi liquid
properties may be modiﬁed as a result of a non-perturbative anomalous dimension for the
fermion field \cite{tesanovic.02}. The preliminary results extracted from fits to a finite volume
equation of state are consistent with a second order phase transition scenario at $N_{fc} \approx 1.5$.
However, as we mentioned in the previous section lattice discretization artifacts are not negligible for 
$N_f \leq 0.5$. This implies that $N_{fc}$ could be even smaller than $1.5$. We are performing simulations 
closer to the continuum limit to clarify this issue. 

\section*{Acknowledgements}
Discussions with Simon Hands and Pavlos Vranas are greatly appreciated.

\end{document}